\documentclass{PoS}

\title{The study of the Three Nucleon Force in full QCD Lattice calculations}

\ShortTitle{The study of the Three Nucleon Force in full QCD Lattice calculations}

\author{\speaker{Takumi Doi}%
\\
Graduate School of Pure and Applied Sciences,
University of Tsukuba,
Tsukuba, Ibaraki 305-8571, Japan\\
        E-mail: \email{doi@ribf.riken.jp}}

\author{for HAL QCD Collaboration}

\author{\includegraphics[width=.30\textwidth]{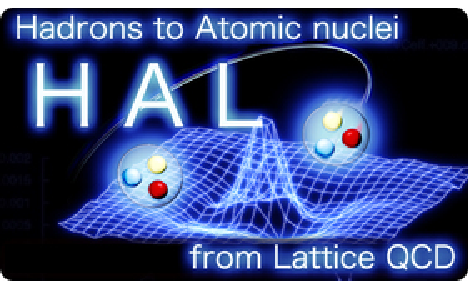}}

\abstract{

We study the three nucleon force in the triton channel
using dynamical clover fermion lattice QCD.
The Nambu-Bethe-Salpeter wave function is utilized
to obtain the potentials among three nucleons.
Since the straightforward calculation is prohibitively expensive,
two different frameworks are developed to meet the challenge.
In the first method,
we study
the effective two nucleon potentials in the three nucleon system,
where the differences between the effective two nucleon potentials 
and the genuine two nucleon potentials 
correspond to the three nucleon system effect,
part of which is originated from the three nucleon force.
The calculation is performed using 
$N_f=2$ clover fermion at $m_\pi= 1.13$ GeV 
generated by CP-PACS Collaboration,
and $N_f=2+1$ clover fermion at $m_\pi= 0.70, 0.57$ GeV
generated by PACS-CS Collaboration.
In the second method,
we study the three nucleon system
with 3D-configuration of nucleons fixed.
This enables us to extract the three nucleon force directly,
if both of parity-even and parity-odd two nucleon potentials are provided.
Since parity-odd two nucleon potentials are not available in lattice QCD
at this moment, 
we propose a new general procedure to identify the 
three nucleon force using only parity-even two nucleon potentials.
The calculation are performed 
with $N_f=2$ clover fermion at $m_\pi= 1.13$ GeV
generated by CP-PACS Collaboration,
employing the linear setup for the 3D-configuration.
Preliminary results for the scalar/isoscalar three nucleon force
are presented.

}

\FullConference{The XXVIII International Symposium on Lattice Field Theory, Lattice2010\\
		June 14-19, 2010\\
		Villasimius, Italy}

\newcommand{\Pu}{p_{\uparrow}}

\newcommand{\Nu}{n_{\uparrow}}
\newcommand{\Nd}{n_{\downarrow}}

\begin{document}

\section{Introduction}
\label{sec:intro}

Since the celebrated work by Yukawa 75 years ago,
the nuclear force has been one of the most essential quantities
in the development of nuclear physics,
and 
so-called ``realistic nuclear potentials''
between two nucleons (2N) are available 
to date.
%
%
%
%
%
%
%
%
%
However, 
recent precise calculations of
few-nucleon systems
clearly point that the 2N force alone is insufficient
to understand the nuclei, 
which calls for three (and/or more) nucleon forces.
Actually, the three nucleon force (TNF) is
projected to play an important and nontrivial role in various phenomena
in nuclear and astro physics.
For the binding energies of light nuclei, 
attractive TNF is required to reproduce the experimental data.
On the other hand,
repulsive TNF is necessary to
reproduce the empirical saturation density of 
symmetric nuclear matter. 
For the EoS 
of asymmetric nuclear matter,
repulsive TNF is required
to 
explain 
the observed 
maximum neutron star mass.
%
%
Recently, 
it is argued that TNF is responsible for 
the anomaly in the drip line and the nontrivial 
magic number of neutron-rich nuclei~\cite{otsuka:tnf}.

Although the experimental/theoretical scrutiny of 
three nucleon scattering
are shedding light on the natures of TNF~\cite{sekiguchi:pd-breakup},
our knowledge on TNF is still quite limited.
Pioneered by Fujita-Miyazawa~\cite{FM},
TNF have been mainly studied from the two-pion exchange picture
with the $\Delta$-excitation.
In addition, repulsive TNF is often introduced 
phenomenologically~\cite{UIX}.
%
Recently, the TNF based on chiral EFT is developing~\cite{UvK:eft},
but the unknown low-energy constants can be obtained 
only by the fitting to the experimental data.
%
%
Since TNF is originated by the fact that
the nucleon is not a fundamental particle,
it is essential to study TNF
from the fundamental DoF, i.e., quarks and gluons.
In this proceeding, 
we report such first-principle calculations of TNF
using lattice QCD.


In the lattice QCD calculation of nuclear forces,
it is recently proposed~\cite{ishii:nf:prl} to use
the Nambu-Bethe-Salpeter (NBS) wave function 
so that the 
potential is faithful to the phase shift
by construction.
The obtained 2N potentials 
are found to have desirable features,
such as attractive well at long and medium
distances, and the central repulsive core at short distance%
~\cite{ishii:nf:prl,ishii:nf:method}.
The method has been successfully extended to the
hyperon-nucleon (YN) and hyperon-hyperon (YY) interactions%
~\cite{nemura:LambdaN,inoue:su3}.

We extend this methodology to the three nucleon (3N) system,
namely, in the triton channel.
Due to the significantly enlarged DoF,
the straightforward calculation is impossible.
We explore two different methods to overcome this problem,
one is the study of effective 2N potentials in 3N system,
and the other is the study of 3N system 
with fixed 3D-configuration by the linear setup.


\section{Formulation for the effective 2N potential}
\label{sec:formulation:eff_2N}

Since the detailed formulation to study a 2N system is 
given in Ref.~\cite{ishii:nf:method},
we discuss the extension to a 3N system here.
We first consider the equal-time NBS wave 
function $\psi(\vec{r},\vec{\rho})$,
which can be obtained by the calculation of six point correlation function,
\begin{eqnarray}
G_{\alpha\beta\gamma,\alpha'\beta'\gamma'} (\vec{r},\vec{\rho},t-t_0)
= \langle\
           N_\alpha(\vec{x}_1,t) N_\beta (\vec{x}_2,t) N_\gamma (\vec{x}_3,t) \
\overline{(N_{\alpha'}'(t_0)       N_{\beta'}'(t_0)        N_{\gamma'}'(t_0))}
\ \rangle ,
\end{eqnarray}
where
$\vec{r} \equiv \vec{x}_1 - \vec{x}_2$, 
$\vec{\rho} \equiv \vec{x}_3 - (\vec{x}_1 + \vec{x}_2)/2$
are the Jacobi coordinates,
and $N$ denotes 
either of 
$p$ or $n$.

At the leading order 
of the velocity expansion of the potentials,
the NBS wave function can be converted to the potentials
through the following 
Schr\"odinger equation,
\begin{eqnarray}
%
\biggl[ 
- \frac{1}{2\mu_r} \nabla^2_{r} - \frac{1}{2\mu_\rho} \nabla^2_{\rho} 
+ \sum_{i<j} V_{2N,ij} (\vec{r}_{ij})
+ V_{TNF} (\vec{r}, \vec{\rho})
\biggr] \psi(\vec{r}, \vec{\rho})
= E \psi(\vec{r}, \vec{\rho}) , 
\label{eq:Sch_3N}
\end{eqnarray}
where
$V_{2N,ij}(\vec{r}_{ij})$ with $\vec{r}_{ij} \equiv \vec{x}_i - \vec{x}_j$
denotes the potential between $(i,j)$-pair,
$V_{TNF}(\vec{r},\vec{\rho})$ the TNF,
$\mu_r = m_N/2$, $\mu_\rho = 2m_N/3$ the reduced masses.
%
%
If we can calculate 
$\psi(\vec{r}, \vec{\rho})$ for all $\vec{r}$, $\vec{\rho}$,
and if all $V_{2N,ij}(\vec{r}_{ij})$ are available 
by (separate) lattice calculations for the genuine 2N system,
we can extract $V_{TNF}(\vec{r},\vec{\rho})$.
Unfortunately, this is not the case:
Since both $\vec{r}$ and $\vec{\rho}$ have $L^3$ DoF,
the calculation cost is more expensive by a factor of $L^3$
compared to the 2N system.
Furthermore, the number of diagrams 
to be calculated in the Wick contraction
tends to diverge with a factor of $N_u ! \times N_d !$
($N_{u,d}$ are numbers of u,d quarks in the system).
We also note that
not all 2N potentials are available in lattice QCD at this moment:
Only parity-even 2N potentials have been obtained so far.

In order to avoid these problems,
we consider the effective 2N potential 
in the 3N system.
More specifically,
we take the summation over the 
location of the spectator nucleon $N(\vec{x}_3)$,
\begin{eqnarray}
\phi(\vec{r}) \equiv 
\sum_{\vec{x}_3} \psi(\vec{r},\vec{\rho}) = 
\sum_{\vec{\rho}} \psi(\vec{r},\vec{\rho}),
\end{eqnarray}
and define the effective potential 
between $N(\vec{x}_1)$ and $N(\vec{x}_2)$
via
the 
effective Schr\"odinger equation,
\begin{eqnarray}
\biggl[ 
- \frac{1}{2\mu_r} \nabla^2_{r}
+ V_{eff} (\vec{r})
\biggr] \phi(\vec{r})
= E \phi(\vec{r}) .
\end{eqnarray}
%

In this calculation,
the DoF of $\vec{\rho}$ is integrated out beforehand,
and thus
the calculation cost is reduced by a factor of 
$\sim 1/L^3$, compared to the straightforward calculation.
Yet, the calculation remains quite expensive
due to the large numbers of Wick contractions,
and
we use several techniques to reduce the calculation cost, 
e.g., we take advantage of symmetries (such as isospin symmetry),
and
we employ the non-relativistic limit for the source nucleon operator.
Similar techniques are
(independently) developed 
in the calculations of binding 
energies of triton/helium nuclei~\cite{yamazaki:He}.

As the 3N system, we study the triton channel,
$I=1/2$, $J^P= 1/2^+$.
Because the spectator nucleon is projected to the S-wave,
the possible quantum numbers between the (effective) 2N
are only $^{2S+1}L_J =$ $^1S_0$, $^3S_1$, $^3D_1$,
and 
we can obtain the effective 2N potentials $V_{eff} (\vec{r})$
in parity-even channel, i.e., 
the central $V_{C,eff}^{I=1,S=0}$,
$V_{C,eff}^{I=0,S=1}$
and the tensor $V_{T,eff}^{I=0,S=1}$
potentials.
%
%
We calculate all counterparts of these parity-even potentials 
in the genuine 2N system, $V_{2N}(\vec{r})$,
and compare them with $V_{eff} (\vec{r})$
to extract the effect of the 3N system.
In one sense, $V_{eff} (\vec{r}) - V_{2N}(\vec{r})$
can be considered to be the ``finite density effect'' in the 3N system.
Some of this effect are attributed 
to the genuine 2N potential with the nontrivial 3N correlation,
and the others are originated by the genuine TNF.
In this way, we can (indirectly) access the effect of TNF.

\section{Lattice setup and the results for the effective 2N potential}
\label{sec:results:eff_2N}

We employ 
$N_f=2$ dynamical 
configurations
with mean field improved clover fermion 
and 
RG-improved gauge action
generated by CP-PACS Collaboration~\cite{conf:cp-pacs}.
We use 
598 configurations at
$\beta=1.95$
with the lattice size of $L^3 \times T = 16^3\times 32$,
which corresponds to
$(2.5{\rm fm})^3$ box in physical spacial size
with 
the lattice spacing of 
$a^{-1} = 1.269 {\rm GeV}$. 
We calculate at the hopping parameter of $u$, $d$ quarks
$\kappa_{ud} = 0.13750$, which corresponds to 
$m_\pi = 1.13$ GeV, $m_N = 2.15$ GeV.
We use the wall quark source with Coulomb gauge fixing.
In order to enhance the statistics, we perform 
the calculation for sources on 16 time slices
for each configuration.

\begin{figure}[bt]
\vspace*{-15mm}
\begin{minipage}{0.45\textwidth}
\begin{center}
\hspace*{-10mm}
\includegraphics[width=0.85\textwidth,angle=270]{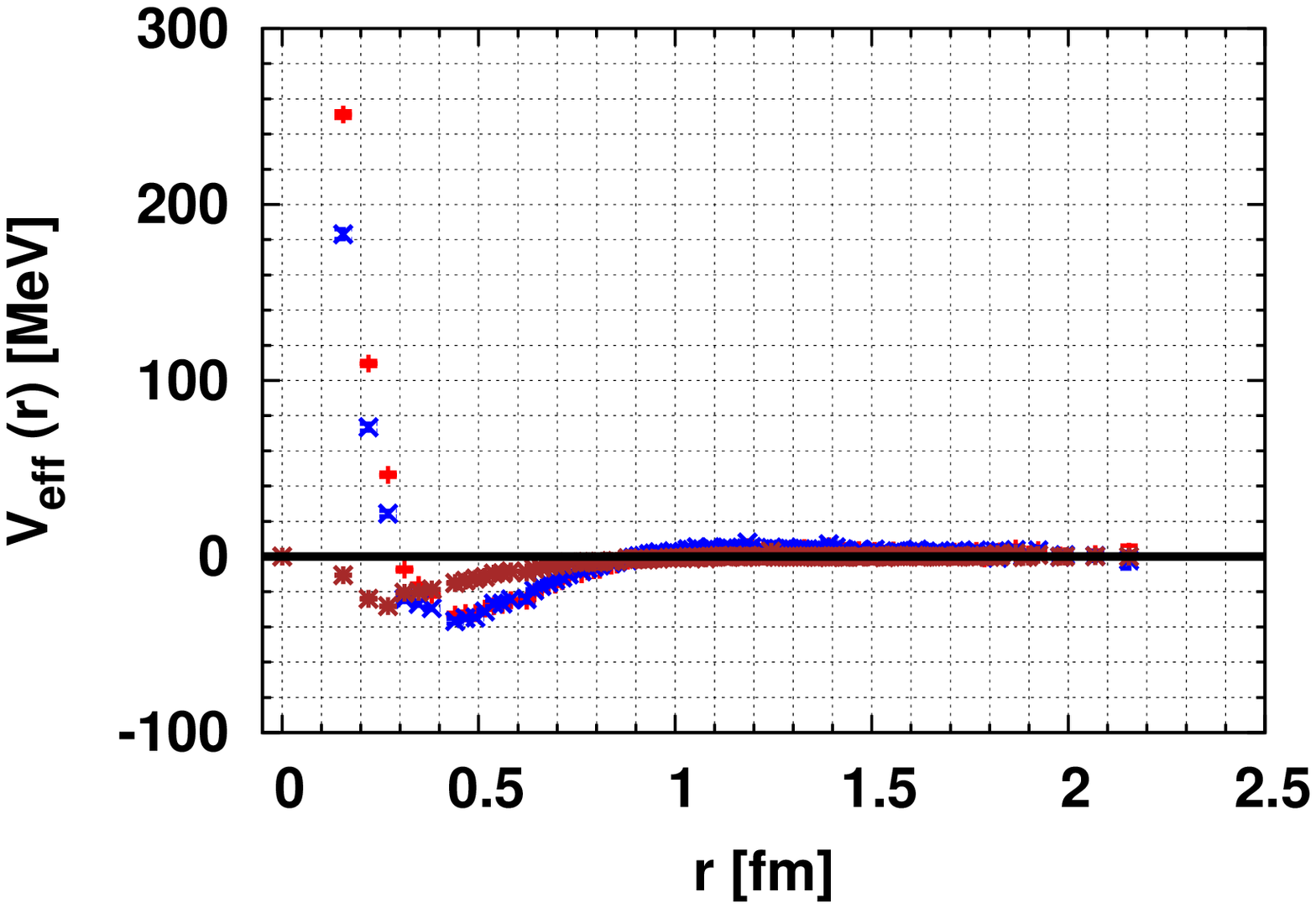}
\caption[hoge]{
\label{fig:pot_3b.eff_2N}
Effective 2N potentials,
where 
red, blue, brown points correspond to 
$V_{C,eff}^{I=1,S=0}$,
$V_{C,eff}^{I=0,S=1}$,
$V_{T,eff}^{I=0,S=1}$
potential, 
respectively.
}
\end{center}
\end{minipage}
\hfill
\begin{minipage}{0.45\textwidth}
\begin{center}
\vspace*{-4mm}
\hspace*{-10mm}
\includegraphics[width=0.85\textwidth,angle=270]{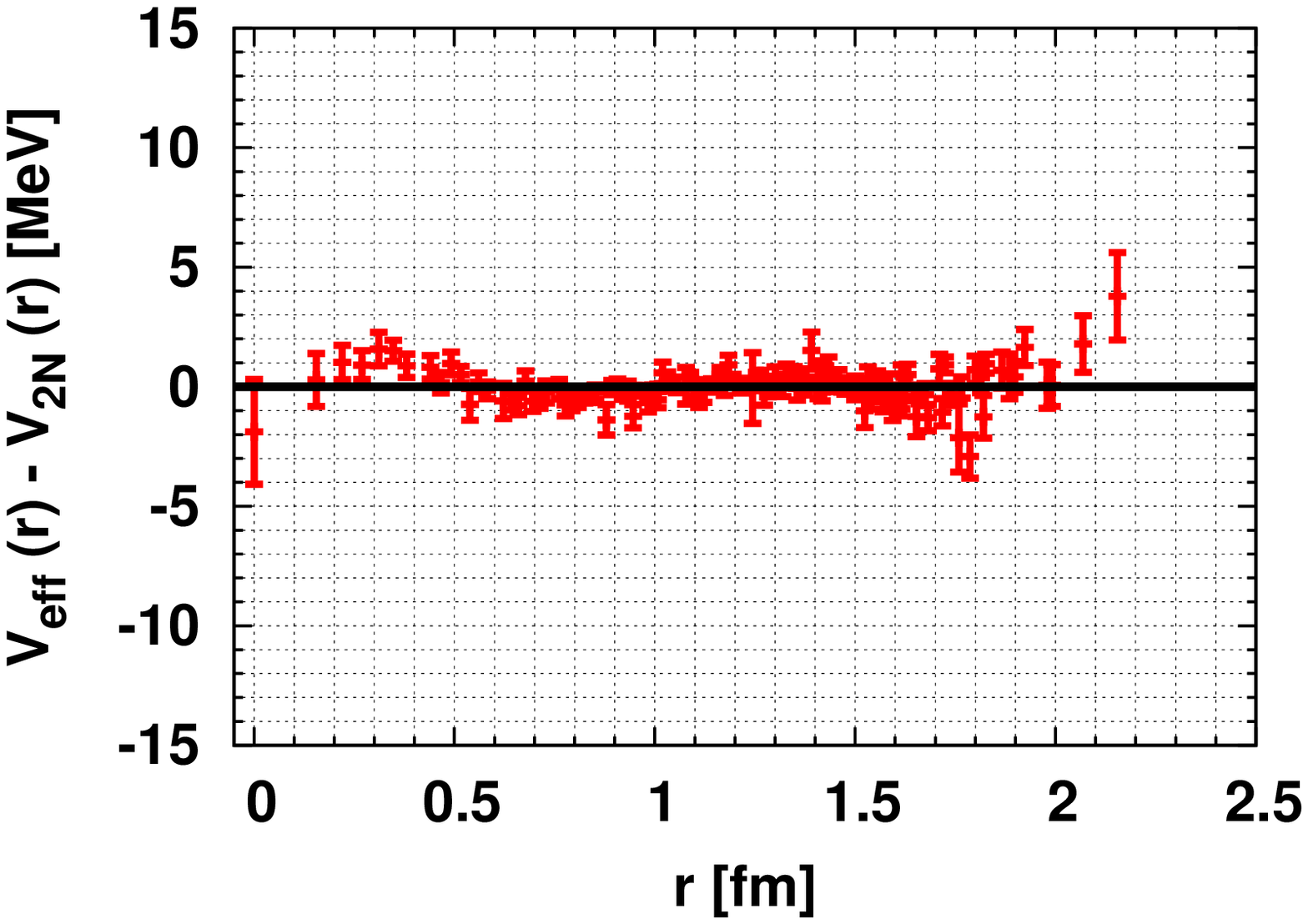}
\caption[hoge]{
\label{fig:pot_diff.1S0}
The difference between the effective 2N and the genuine 2N
for $V_{C}^{I=1,S=0}$ potential.
}
\end{center}
\end{minipage}
\end{figure}

\begin{figure}[bt]
\begin{minipage}{0.45\textwidth}
\begin{center}
\vspace*{-12mm}
\hspace*{-10mm}
\includegraphics[width=0.85\textwidth,angle=270]{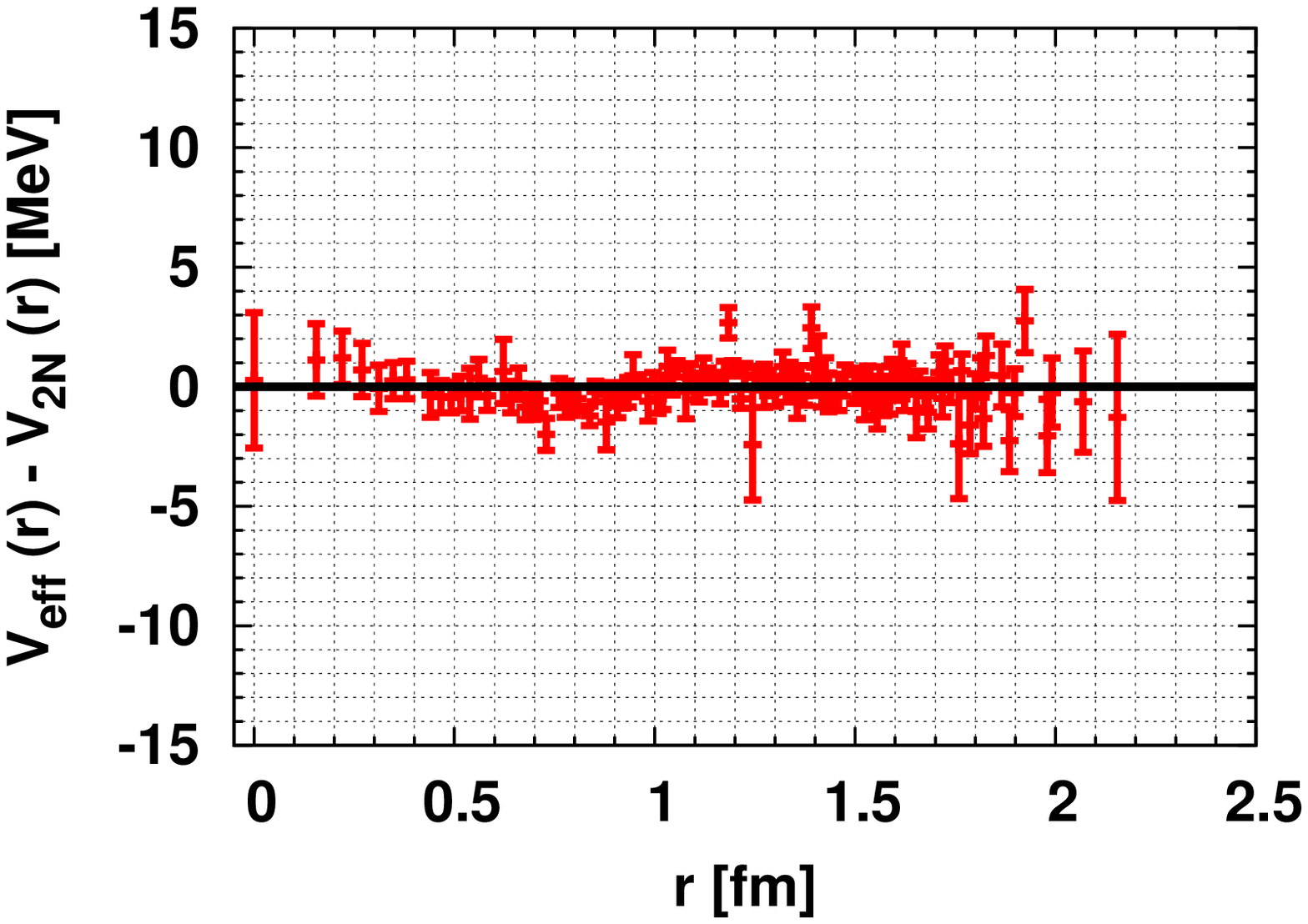}
\caption[hoge]{
\label{fig:pot_diff.cen}
Same as Fig.~\ref{fig:pot_diff.1S0},
but for $V_{C}^{I=0,S=1}$. 
}
\end{center}
\end{minipage}
\hfill
\begin{minipage}{0.45\textwidth}
\begin{center}
\vspace*{-12mm}
\hspace*{-10mm}
\includegraphics[width=0.85\textwidth,angle=270]{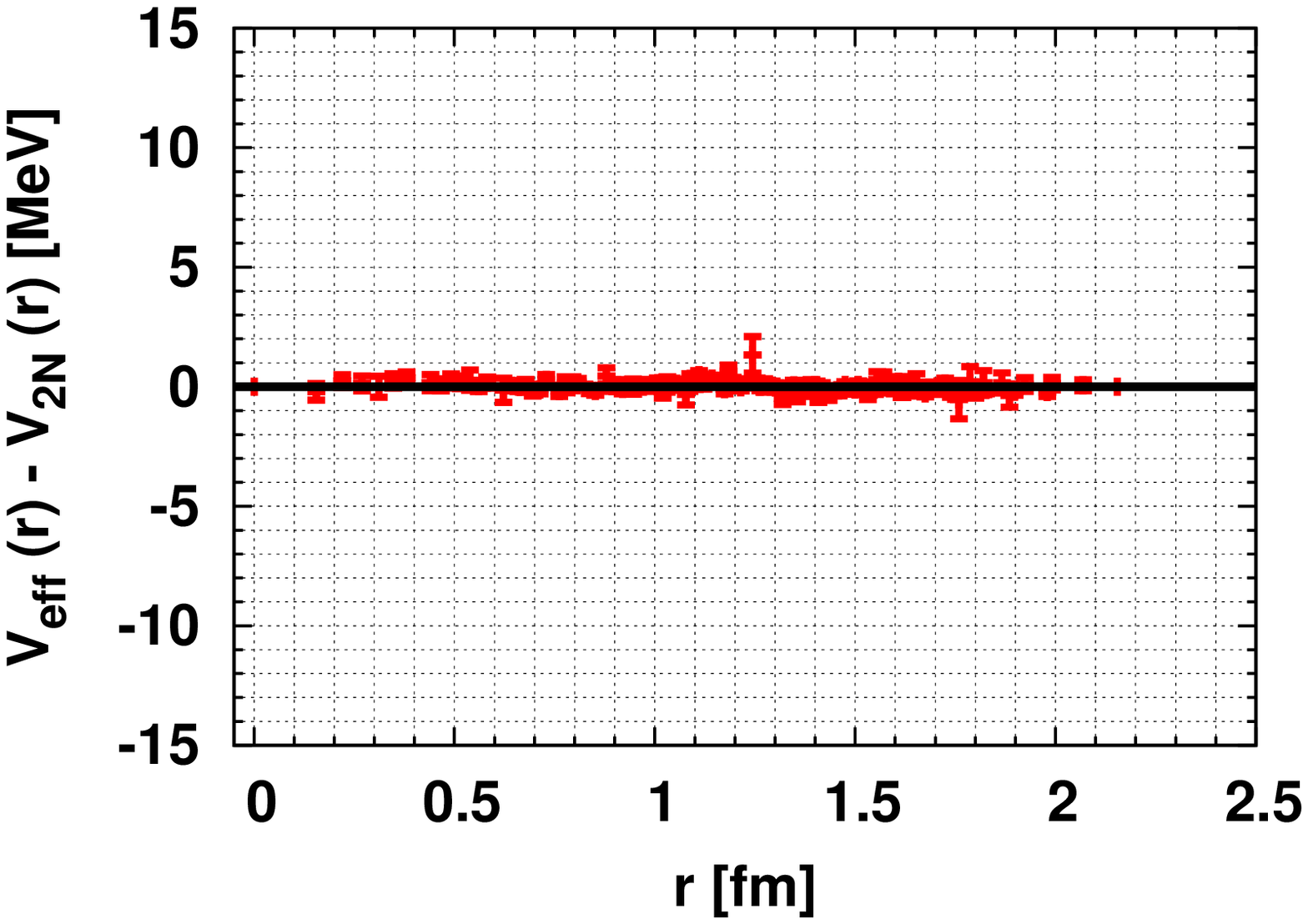}
\caption[hoge]{
\label{fig:pot_diff.ten}
Same as Fig.~\ref{fig:pot_diff.1S0},
but for $V_{T}^{I=0,S=1}$. 
}
\end{center}
\end{minipage}
\vspace*{-0mm}
\end{figure}

In Fig.~\ref{fig:pot_3b.eff_2N}, 
we show preliminary results for 
$V_{eff}(r)$
in the triton channel at $t-t_0=8$.
Here, the constant shift by energy is not included
for the central potentials.
What is noteworthy is that 
$V_{eff}(r)$
are obtained with good precision.
This is quite nontrivial,
since the S/N usually gets worse
for more quarks in the system.
In Figs.~\ref{fig:pot_diff.1S0},\ref{fig:pot_diff.cen},\ref{fig:pot_diff.ten},
we plot $V_{eff}(r) - V_{2N}(r)$ 
for each potential.
One can see that the discrepancy is 
consistent with zero within error-bar.
In particular, the tensor potential is best constrained
within several MeV statistical error,
and there is no indication of the TNF effect.

One of the possible explanations why the TNF effect is not observed
is that 
the TNF is suppressed at heavy quark mass.
Actually, pion exchange is expected to be strongly suppressed
with $m_\pi = 1.13$ GeV in this calculation.
Therefore, we investigate the quark mass dependence
using the configurations with smaller quark masses,
generated by PACS-CS Collaboration~\cite{conf:pacs-cs}:
$N_f = 2+1$ nonperturbatively ${\cal O}(a)$ improved clover fermion, 
$V = 32^3 \times 64$, 
$\beta = 1.90$, 
$a^{-1} = 2.18 {\rm GeV}$.
We use 399 configurations at $\kappa_{ud} = 0.13700$, $\kappa_{s} = 0.13640$
 ($m_\pi = 0.70$ GeV, $m_N = 1.58$ GeV)
and 400 configurations at $\kappa_{ud} = 0.13727$, $\kappa_{s} = 0.13640$
($m_\pi = 0.57$ GeV, $m_N = 1.41$ GeV).
For both quark mass setups,
we perform 4 source time slice measurements for each configuration.

We obtain basically similar results to previous results:
For all potentials at either quark mass, 
the differences between $V_{eff}(r)$ and $V_{2N}(r)$
are found to be consistent with zero within error-bar.
In particular, the statistical error of $V_{eff}(r)-V_{2N}(r)$ in tensor potential
remains less than 5 MeV even for the lightest quark mass.
This may indicate the necessity to decrease the quark mass further.

Another possible explanation is that
the TNF effect is obscured by the summation over the location of 
the spectator nucleon.
In fact, while the TNF effect is expected to be
enhanced when all three nucleons are close to each other,
such 3D-spacial configurations make small contributions in the
spectator summation.
In order to assess this possibility, 
we study the 3N system with (at closely) fixed 3D-configuration,
which is the subject of the next section.


\section{Formulation for the fixed 3D-configuration with the linear setup}
\label{sec:formulation:linear}

We consider Eq.~(\ref{eq:Sch_3N}) with fixed 3D-configuration of $\vec{r}$, $\vec{\rho}$.
The advantage of this method is that
the subtraction procedure of the genuine 2N potentials $V_{2N}$ becomes clearer in principle,
and thus the TNF can be extracted directly.
Furthermore, by choosing small $|\vec{r}|$, $|\vec{\rho}|$ configurations,
the TNF effect is expected to be enhanced.
The disadvantage of this calculation is the calculation cost:
It is more expensive by a factor of ${\cal O}(10)$-${\cal O}(10^2)$,
compared to the effective 2N potential study.
%
%
Therefore, we restrict the calculation to limited 3D-configurations.
In addition,
efforts are taken to speed up the calculation code.
As the fixed 3D-configuration, we take the linear setup with $\vec{\rho}=\vec{0}$.
Just for the sake of convenience,
we redefine as $\vec{r} \rightarrow 2\vec{r}$ hereafter.
In other words,
three nucleons are aligned linearly with equal spacings of $r=|\vec{r}|$ in this linear setup.

The advantage of the linear setup is it's simplicity.
Because of $\vec{\rho}=\vec{0}$, the third nucleon is attached
to $(1,2)$-nucleon pair with only S-wave.
Considering the total 3N quantum numbers of $I=1/2, J^P=1/2^+$, 
the wave function can be completely spanned by
only three bases, which can be labeled
by the quantum numbers of $(1,2)$-pair as
$^1S_0$, $^3S_1$, $^3D_1$.
Therefore, the Schr\"odinger equation 
can be 
simplified to
the $3\times 3$ coupled channel equations
with the bases of 
$\psi_{^1S_0}$, $\psi_{^3S_1}$, $\psi_{^3D_1}$.
The reduction of the dimension of bases 
is expected to improve the S/N as well.

Unfortunately,
even by the calculation of the fixed 3D-configuration (including the linear setup),
the subtraction of $V_{2N}$ 
remains nontrivial.
As was noted in Sec.~\ref{sec:formulation:eff_2N},
the parity-odd potentials are not available
in lattice QCD at this moment,
and we cannot subtract them unambiguously.
Note that although the total parity of the 3N system can be projected, 
a 2N-pair 
could be either of positive or negative parity.
The familiar procedure of partial wave expansion cannot be performed here,
since we can calculate only limited 3D-configurations due to the huge
calculation cost.

In the effort to overcome this issue,
we find that the following channel in 
the triton is useful,
\begin{eqnarray}
\psi_S &\equiv&
\frac{1}{\sqrt{6}}
\Big[
-   \Pu \Nu \Nd + \Pu \Nd \Nu               
                - \Nu \Nd \Pu + \Nd \Nu \Pu 
+   \Nu \Pu \Nd               - \Nd \Pu \Nu
\Big]  .
\label{eq:psi_S}
\end{eqnarray}
This wave function itself has been well known,
but the point here is that 
it is anti-symmetric
in spin/isospin spaces 
for any 2N-pair.
Combined with the Pauli-principle,
it is automatically guaranteed that
any 2N-pair couples with even parity only.
Therefore, we can extract the TNF unambiguously in this channel,
without the information of parity-odd 2N potentials.
Note that no assumption on the choice of 3D-configuration 
is imposed in this argument,
and we can take advantage of this feature
for the future TNF calculations with 
3D-configurations other than the linear setup.

Coming back to the linear setup,
we examine
the explicit form of the potential matrix of $V_{2N}$.
At the leading order 
of the velocity expansion,
$V_{2N}$ can be
written in terms of
center $V_C^{IS}$ and tensor $V_T^{IS}$ potentials
with isospin $I$ and spin $S$,
$V_C^{00}$, 
$V_C^{10}$, 
$V_C^{01}$, 
$V_C^{11}$, 
$V_T^{01}$,
$V_T^{11}$,
where the label ``$2N$'' is omitted for simplicity.
Explicit calculation gives us

%
%
%
{\scriptsize 
\begin{eqnarray}
\label{eq:pot_3H_sym}
V_{2N}
=
\left(
\renewcommand{\arraystretch}{1.8}
\begin{array}{c|c|c}
                     +             V_C^{10}(r) +             V_C^{01}(r)                    & 
                     + \frac{1}{2} V_C^{10}(r) - \frac{1}{2} V_C^{01}(r)                    & 
- 2 V_T^{01}(r)   \\                                                                            
\qquad + \frac{1}{2} V_C^{10}(2r) + \frac{1}{2} V_C^{01}(2r) & 
\qquad - \frac{1}{2} V_C^{10}(2r) + \frac{1}{2} V_C^{01}(2r) & \qquad\qquad + 2 V_T^{01}(2r)  \\[2mm] \hline 
                        + \frac{1}{2} V_C^{10}(r) - \frac{1}{2} V_C^{01}(r)                          & 
+\frac{3}{4} V_C^{00}(r) + \frac{1}{4} V_C^{10}(r) + \frac{1}{4} V_C^{01}(r) + \frac{3}{4} V_C^{11}(r) & 
+ V_T^{01}(r) - 3 V_T^{11}(r)  \\                                                                
\qquad - \frac{1}{2} V_C^{10}(2r) + \frac{1}{2} V_C^{01}(2r) & 
\qquad + \frac{1}{2} V_C^{10}(2r) + \frac{1}{2} V_C^{01}(2r) & \qquad\qquad + 2 V_T^{01}(2r)  \\[2mm] \hline 
- 2 V_T^{01}(r)   &                                                                       
+ V_T^{01}(r) - 3 V_T^{11}(r)  &                                                           
+ \frac{1}{2} V_C^{01}(r) + \frac{3}{2} V_C^{11}(r) - V_T^{01}(r) - 3 V_T^{11}(r) \\         
\qquad\qquad + 2 V_T^{01}(2r) & \qquad\qquad + 2 V_T^{01}(2r) & \qquad\qquad + V_C^{01}(2r) - 2 V_T^{01}(2r)  
\end{array}
\right)
\end{eqnarray}
} 
Here, we span the spaces with the rotated bases given by
$(\psi_S, \psi_M, \psi_{^3D_1})^T$,
where
$\psi_S$ in Eq.~(\ref{eq:psi_S}) is shown to be
$\psi_S = \frac{1}{\sqrt{2}} ( - \psi_{^1S_0} + \psi_{^3S_1} )$,
and
$\psi_M \equiv \frac{1}{\sqrt{2}} ( + \psi_{^1S_0} + \psi_{^3S_1} )$.
Note that 
neither of 
parity-odd 2N potentials,
$V_C^{00}$, $V_C^{11}$, $V_T^{11}$, appear in the 
first 
row
in Eq.~(\ref{eq:pot_3H_sym}), as was discussed previously.

\vspace*{-1mm}
\section{The lattice QCD results for the linear setup}
\label{sec:results:linear}
\vspace*{-2mm}

We employ the 
CP-PACS $N_f=2$ clover fermion configurations~\cite{conf:cp-pacs},
which are used for the effective 2N potential study
in Sec.~\ref{sec:results:eff_2N}.
598 configurations are used with 16 source time slice measurements
for each configuration.
We perform the calculation
with the linear setup at 7 physical points of the distance $r$.
As is explicitly shown in Eq.~(\ref{eq:pot_3H_sym}),
we have only one channel which is free from parity-odd 2N potentials.
Correspondingly, we can determine one type of TNF.
In this proceeding,
we consider the scalar/isoscalar type TNF.
In fact, in the Urbana IX model~\cite{UIX},
which is an often-used phenomenological TNF,
TNF consists of two parts,
one is the two-pion exchange TNF (which has spin/isospin dependencies),
and the other is phenomenologically introduced 
scalar/isoscalar repulsive TNF.
Since the two-pion exchange TNF is 
expected to be suppressed in the current lattice setup 
of $m_\pi = 1.13$ GeV,
it is reasonable to assume the scalar/isoscalar nature 
for TNF here.

\begin{figure}[t]
\vspace*{-15mm}
\begin{minipage}{0.45\textwidth}
\begin{center}
\hspace*{-8mm}
\includegraphics[width=0.85\textwidth,angle=270]{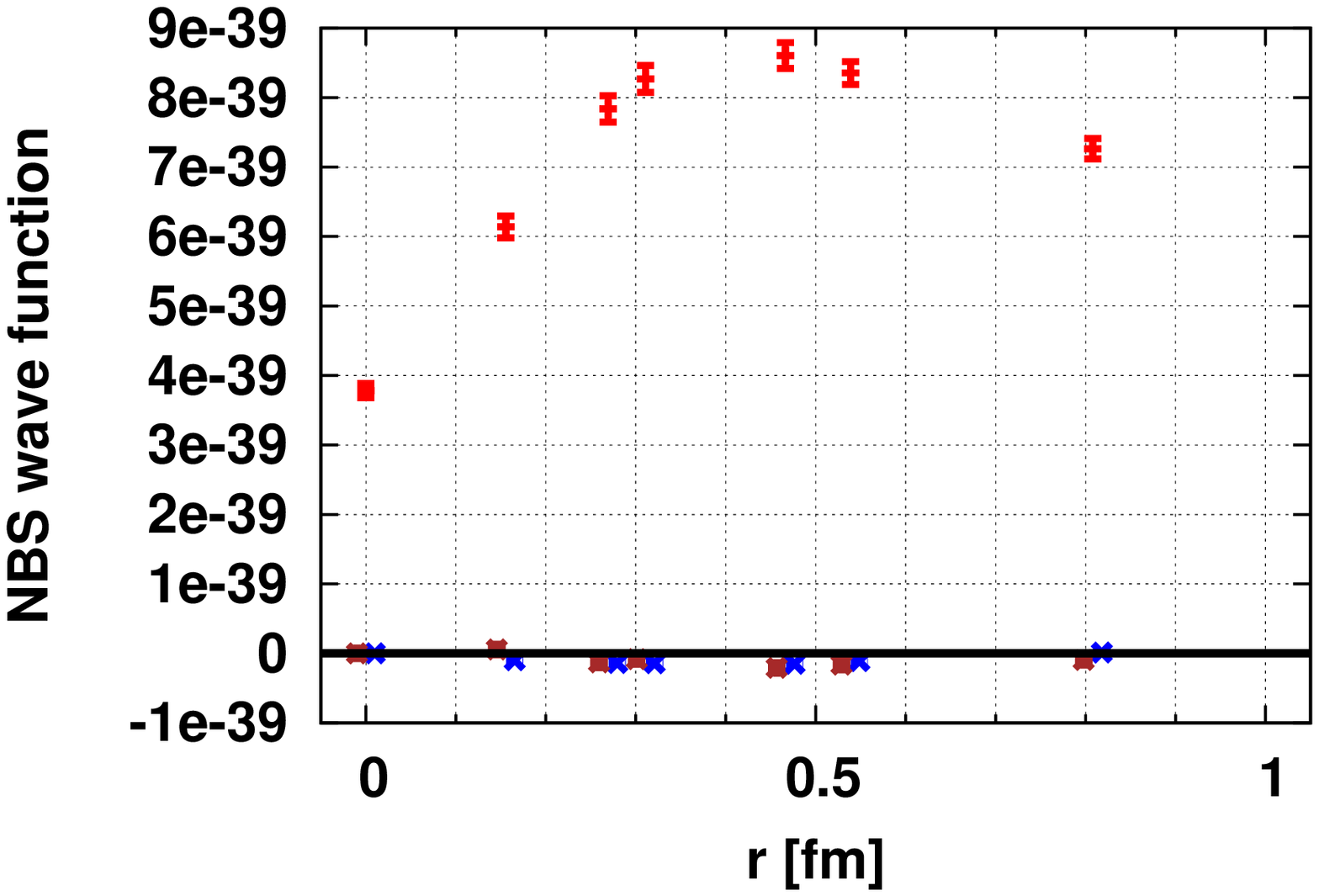}
\caption{
\label{fig:wf}
The wave function with linear setup in the triton channel.
Red, blue, brown points correspond to
$\psi_S$, $\psi_M$, $\psi_{^3D_1}$, respectively.
}
\end{center}
\end{minipage}
\hfill
\begin{minipage}{0.45\textwidth}
\begin{center}
\hspace*{-12mm}
\includegraphics[width=0.85\textwidth,angle=270]{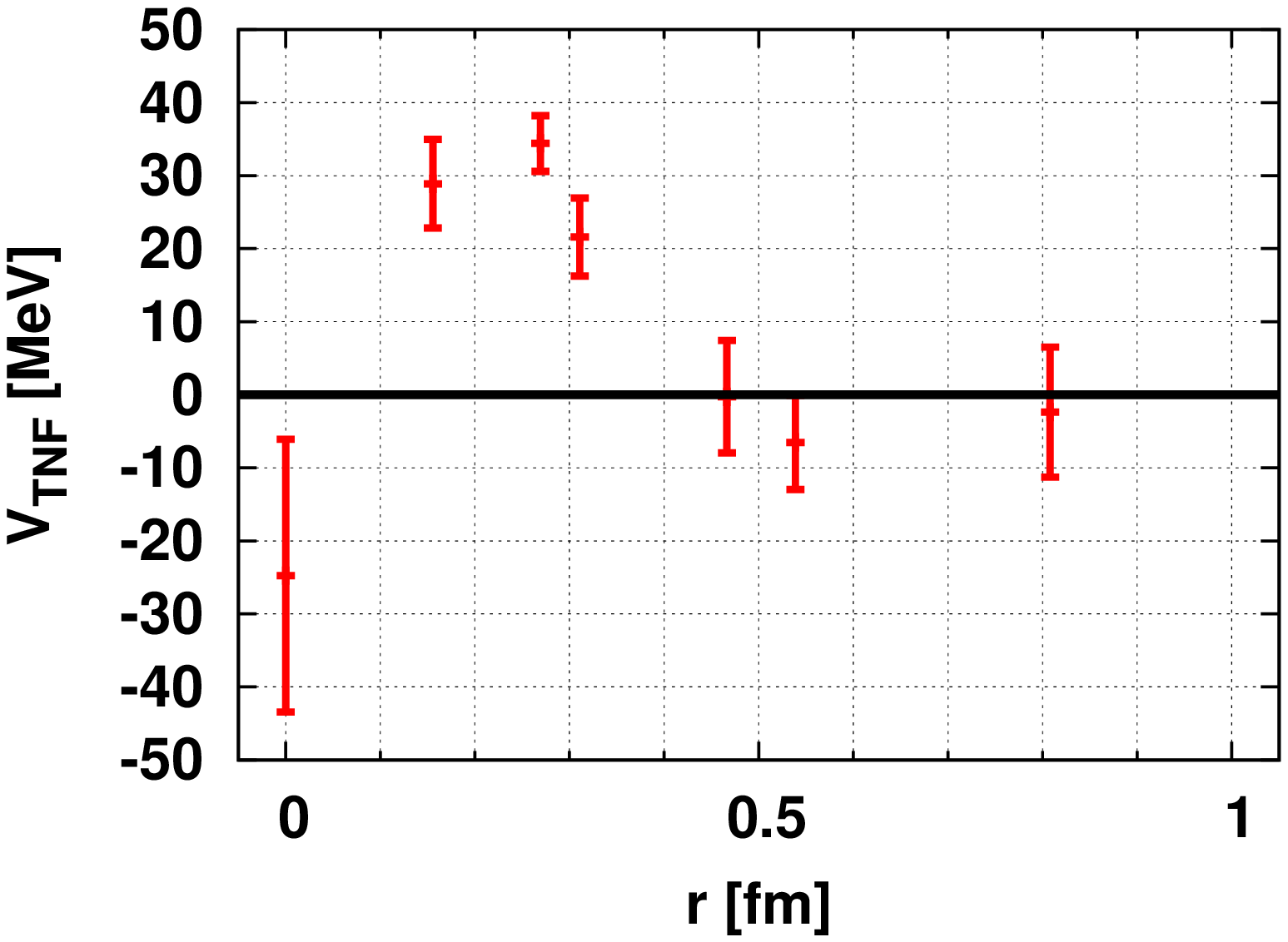}
\caption{
\label{fig:TNR}
The scalar/isoscalar TNF in the triton channel,
plotted against the distance $r$ in the linear setup.
}
\end{center}
\end{minipage}
\vspace*{-2mm}
\end{figure}

In Fig.~\ref{fig:wf},
we plot each wave function of
$\psi_S$, $\psi_M$, $\psi_{^3D_1}$ 
in the triton channel at $t-t_0 = 8$.
We observe that 
$\psi_S$ dominates the wave function.
This is because 
$\psi_S$ contains the component for which
all three nucleons are in S-wave.
We emphasize that the observation of good S/N 
for the wave function is much more nontrivial
than the effective 2N study,
because the practical statistical sampling number becomes much smaller
by fixing the 3D-configuration.

By subtracting the $V_{2N}$ in Eq.~(\ref{eq:pot_3H_sym})
from the total potentials in the 3N system,
we determine the TNF.
In Fig.~\ref{fig:TNR}, we plot the preliminary results
for the scalar/isoscalar TNF. 
Here, the $r$-independent shift by energies is not included,
and thus about ${\cal O}(10)$ MeV systematic error is understood.
There are various physical implications in Fig.~\ref{fig:TNR}.
At the long distance region of $r$, the TNF is small as is 
expected.
At the short distance region, 
we observe the indication of repulsive TNF.
Recalling that the repulsive short-range TNF is phenomenologically required 
to explain the saturation density of nuclear matter, etc.,
this is very encouraging result.
Of course, we note that
further study is necessary to confirm this result,
e.g., the study of the ground state saturation,
the evaluation of the constant shift by energies,
the examination of the discretization error.

\vspace*{-1mm}
\section{Summary}
\vspace*{-2mm}

We have studied the three nucleon force (TNF) in the triton channel
in lattice QCD, developing two different methods.
In the first method, 
effective 2N potentials have been studied
using $N_f=2$ clover fermion at $m_\pi= 1.13$ GeV,
and $N_f=2+1$ clover fermion at $m_\pi= 0.70, 0.57$ GeV.
The effective 2N potentials have been found to be
consistent with genuine 2N potentials within error.
In the second method,
we have fixed the 3D-configuration of three nucleons.
In particular, we have established the 
general procedure which can identify the TNF
without the information of parity-odd 2N potentials.
The calculation have been performed 
with $N_f=2$ clover fermion at $m_\pi= 1.13$ GeV
with the linear setup for the 3D-configuration,
and the indication of repulsive TNF at short distances
have been obtained.
Further work to confirm the results is currently underway.


We thank 
CP-PACS and PACS-CS Collaborations 
and 
ILDG/JLDG~\cite{conf:ildg/jldg} 
for providing the configurations.
TD is supported in part by
Grant-in-Aid for JSPS Fellows 21$\cdot$5985.
This research is supported in part by 
Grant-in-Aid for Scientific Research on Innovative Areas (No.2004:20105001,
20105003)
and the Large Scale Simulation Program No.09-23 (FY2009) of KEK.
The numerical calculations have been performed
on T2K at University of Tsukuba 
and
Blue Gene/L at KEK.

\end{document}